\documentclass[journal]{IEEEtran}

\ifCLASSINFOpdf
\else
   \usepackage[dvips]{graphicx}
\fi
 
\usepackage{multirow}
\usepackage{amsmath}
\usepackage{amsmath,graphicx}
\usepackage{epstopdf}
\usepackage{array}
\usepackage{graphicx}
\usepackage{multirow}
\usepackage{color, soul}
\usepackage{cite}
\usepackage[colorlinks,linkcolor=blue]{hyperref}
\usepackage{cite}
\usepackage{graphicx}
\usepackage{bm}
\newcommand{\tabincell}[2]{\begin{tabular}{@{}#1@{}}#2\end{tabular}}
\usepackage{makecell}

\usepackage{longtable}
\usepackage{xparse}

\begin{document}
\title{Modeling Prosodic Phrasing with Multi-Task Learning in Tacotron-based TTS}

\author{Rui Liu, \IEEEmembership{Member, IEEE},  Berrak Sisman, \IEEEmembership{Member, IEEE}, Feilong Bao, Guanglai Gao,\\ Haizhou Li, \IEEEmembership{Fellow, IEEE}
\thanks{Rui Liu, Feilong Bao and Guanglai Gao are with the Department of Computer Science, Inner Mongolia University. Rui Liu is also with the National University of Singapore. Berrak Sisman is with the Information Systems Technology and Design (ISTD) Pillar at Singapore University of Technology and Design. Haizhou Li is with the Department of Electrical and Computer Engineering, National University of Singapore.}
\thanks{This research is supported by the National Research Foundation, Singapore under its AI Singapore Programme (Award No: AISG-GC-2019-002) and (Award No: AISG-100E-2018-006), and its National Robotics Programme (Grant No. 192 25 00054), and by RIE2020 Advanced Manufacturing and Engineering Programmatic Grants A1687b0033, and A18A2b0046. This research is also supported by SUTD Start-up Grant Artificial Intelligence for Human Voice Conversion (SRG ISTD 2020 158), SUTD  AI Grant 'The Understanding and Synthesis of Expressive Speech by AI' (PIE-SGP-AI-2020-02) and China National Natural Science Foundation (No.61773224).}
}

\maketitle

\vspace{-5mm}
\begin{abstract}
Tacotron-based end-to-end speech synthesis has shown remarkable voice quality. However, the rendering of prosody in the synthesized speech remains to be improved, especially for long sentences, where prosodic phrasing errors can occur frequently. In this paper, we extend the Tacotron-based speech synthesis framework to explicitly model the prosodic phrase breaks. We propose a multi-task learning scheme for Tacotron training, that optimizes the system to predict both Mel spectrum and phrase breaks.  To our best knowledge, this is the first implementation of multi-task learning for Tacotron based TTS with a prosodic phrasing model. Experiments show that our proposed training scheme consistently improves the voice quality for both Chinese and Mongolian systems.

\end{abstract}


\IEEEpeerreviewmaketitle

\vspace{-4mm}
\section{Introduction}
With the advent of deep learning, end-to-end text-to-speech (TTS) has shown many advantages over the conventional TTS techniques \cite{tokuda2013speech, ze2013statistical}. Tacotron-based approaches \cite{wang2017tacotron,shen2018natural,lee2019robust, ICASSP2020liu, liu2020wavetts} with an encoder-decoder architecture and attention mechanism have shown remarkable performance. The key idea is to integrate the conventional TTS pipeline into a unified network and learn the mapping directly from the text-waveform pair \cite{chung2019semi,He2019,Luong2019}. The recent progress in neural vocoder \cite{hayashi2017investigation,shen2018natural,Okamoto2019, berrak_is18, berrak-journal, sisman2018adaptive} also contributes to the improvement of speech quality. 

Speech prosody includes affective prosody and linguistic prosody. Affective prosody represents the emotion of a speaker, while linguistic prosody relates to the language content. They are both crucial in speech communication. A TTS system is expected to synthesize the right prosodic pattern at the right time. However, most of the current end-to-end systems \cite{wang2017tacotron,shen2018natural,liu2020wavetts, ICASSP2020liu} have not explicitly modeled speech prosody. Therefore, they can't control well the melodic and rhythmic aspects of the generated speech. This usually leads to monotonous speech, even when models are trained on very expressive speech datasets. In this paper, we would like to study the way to enable Tacotron-based TTS  for expressive prosody generation.
 
Multi-task learning (MTL) is a learning paradigm that leverages information from multiple related tasks to help improve the overall performance \cite{zhang2017survey}. MTL is inspired by human learning activities where people often apply the knowledge learned from many tasks for learning a new task, that is called inductive transfer. For example, if we learn to read and write together, the experience in reading can strengthen the writing and vice versa.
MTL has been widely used in speech enhancement \cite{chen2015speech}, and speech recognition \cite{kim2017joint}. It has also been used in speech synthesis \cite{hu2015fusion}, such as statistical parametric speech synthesis with GANs \cite{yang2017statistical} and DNN-based speech synthesis with stacked bottleneck features~\cite{wu2015deep}. In this paper, we apply multi-task learning to the Tacotron-based TTS for prosody modeling.

\begin{figure*}[t]
  \centering
   \vspace{-4mm}
  \includegraphics[width=0.98 \textwidth]{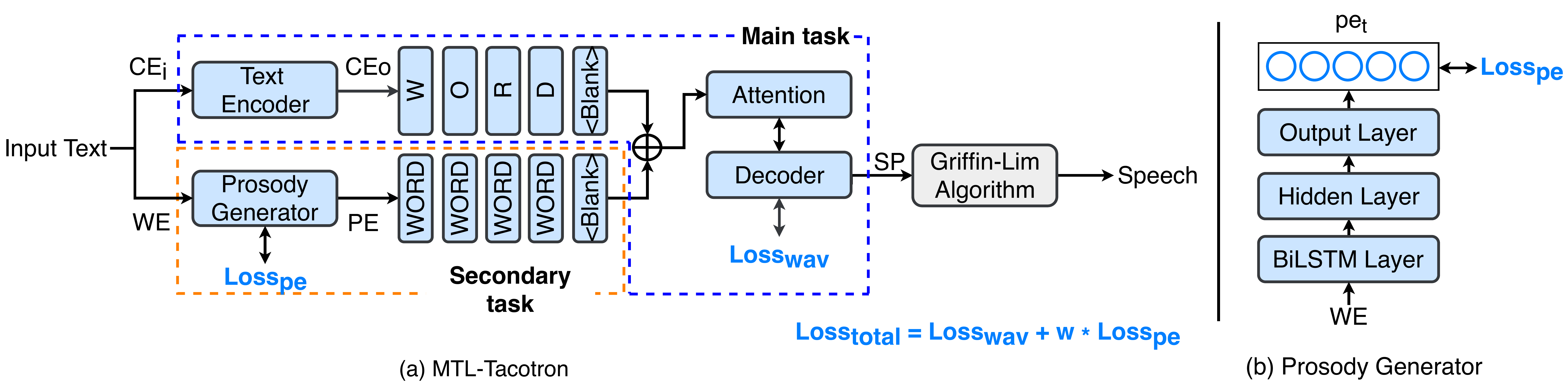}\\
 \vspace{-4mm}
  \caption{Block diagrams of the proposed (a) MTL-Tacotron, and (b) prosody generator. MTL-Tacotron employs a prosody generator to explicitly model prosodic phrasing. The prosody generator produces prosody embedding ($P\!E$) from input text, that forms a joint embedding vector with character embedding. Griffin-Lim algorithm is not involved in training. SP denotes mel-spectrum speech features. $pe_t$ is a 5-dimension prosody embedding. } 
 \vspace{-6mm}
  \label{model_fig}
\end{figure*}

The study on expressive speech synthesis is focused on prosody modeling \cite{jauk2018expressive,akuzawa2018expressive,Mass2018Word}, where speech prosody generally refers to intonation, stress, speaking rate, and phrase breaks. Prosodic phrasing \cite{pb1,pb2,taylor1998assigning,mishra2015intonational} plays an important role in both affective and linguistic expressions. Inadequate phrase breaks may lead to misperception in speech communication. There have been recent studies on prosody modeling for end-to-end TTS system \cite{sloanprosody}, for example, to improve the prosodic phrasing  by using contextual information ~\cite{lu2019implementing}, and syntactic features \cite{guo2019exploiting}. They are incorporated in the stage of text preprocessing, therefore, there are not optimized as part of the synthesis processing.

We propose a novel two-task learning scheme for Tacotron-based TTS model to improve the prosodic phrasing: 1) the main task learns the prediction of the speech spectrum parameters from character-level embedding representation, and 2) the secondary task learns the prediction of a word-level prosody embedding. During training, the secondary task serves as an additional supervision for Tacotron to learn the exquisite prosody structure associated with the input text. At run-time, the prosody embedding serves as a local condition that controls the prosodic phrasing during voice generation.

The main contributions of this paper include: 1) a novel Tacotron-based TTS architecture that explicitly models prosodic phrasing; and 2) a multi-task learning scheme, that
optimizes the model for high quality speech spectrum, and adequate prosodic phrasing at the same time. The proposed system achieves remarkable voice quality for both Chinese Mandarin and Mongolian. To our best knowledge, this is the first multi-task Tacotron implementation that includes an explicit prosodic model.

This paper is organized as follows. Section \ref{sec:baseline} recaps the Tacotron TTS framework. We propose the  multi-task Tacotron in Section~\ref{sec:model} and report the experiments in Section~\ref{exp}. Section \ref{con} concludes the discussion.

\vspace{-3mm} 
\section{Tacotron-based TTS}
\label{sec:baseline}
\vspace{-1mm} 

Tacotron \cite{wang2017tacotron} is a sequence-to-sequence speech synthesizer that consists of an encoder, and a decoder with attention mechanism. 
The encoder takes a sequence of text characters as input, where each character is encoded as a one-hot vector and embedded into a continuous vector, that is called input character embedding. 
The encoder is trained as part of Tacotron to take the input character embedding and generates the output character embedding.
The decoder is an autoregressive recurrent neural network that converts the character embeddings into a sequence of Mel spectrum feature vectors with an attention mechanism.

Just like most of other TTS systems, Tacotron  \cite{wang2017tacotron} is trained to predict the Mel spectrum features from input sequence of characters. Prosody, if taken into consideration, is modeled from the statistics of the training data \cite{wang2017tacotron,chung2019semi,shen2018natural,ICASSP2020liu}. We note that the character sequences themselves are not the most suitable for describing prosody. They do not generalize well because prosody is manifested over a speech segment beyond characters and phonemes. There have been attempts \cite{chung2019semi,ming2019feature} to use word embedding as input to improve the expressiveness of Tacotron-based TTS model, that shows word embedding is prosody-informing.

Another idea~ \cite{wang2018style,Stanton2018Predicting,skerry2018towards,yasuda2019investigation,sun2020generating} is to extract the latent prosody embeddings to characterize prosody. Some~\cite{wang2018style,Stanton2018Predicting,skerry2018towards} learn speech variations without explicit annotations for prosody or style. The learned prosody embeddings are usually not fully controllable and interpretable.  Others~\cite{yasuda2019investigation,sun2020generating} just take the prosody embeddings as an auxiliary input to the TTS model. 

In this paper, we propose a novel prosody embedding, that directly interprets the phrase breaks from the input text. We also propose a novel multi-task learning framework that optimizes the system to generate Mel spectrum, at the same time, accurately predict phrase breaks, which will be the focus of
Section \ref{sec:model}.

\vspace{-4mm}
\section{Tacotron with Multi-Task Learning}
\label{sec:model}
\vspace{-1mm}

We propose multi-task learning \cite{Caruana1993Multitask} for Tacotron as illustrated in Fig. \ref{model_fig}(a), that is referred to as \textit{MTL-Tacotron}. The idea is to dedicate a prosody modeling task to model the prosodic phrasing, that is trainable from data. In the multi-task learning, not only do we optimize the output speech quality, but we also ensure that Tacotron is optimized to produce adequate phrase breaks. We study a two-task learning strategy, 1) the \textit{main task} generates the Mel-spectrums from the input character sequence; and 2) the \textit{secondary task} predicts an appropriate prosodic phrasing.  
 
\vspace{-4mm}
\subsection{Main Task: Spectral Modeling}
\vspace{-1mm}
\label{maintask}
The main task has a network architecture identical to the traditional Tacotron \cite{wang2017tacotron} as shown in Fig. \ref{model_fig}(a). It contains a text encoder and a decoder with attention mechanism. We first convert the word sequence in raw input text to input character embeddings, denoted as $C\!E_i$, which are encoded into output character embeddings, denoted as $C\!E_o$, from which the decoder generates Mel spectral features.

The main task is optimized using a Mel spectral loss function, ${Loss_{wav}} = \sum_{t=1}^{T'} L_{2} (y_{t},y'_{t})$, where $y_{t}$ and $y'_{t}$ are the target and the predicted Mel-spectrum respectively, $T'$ is the total number of Mel spectrum features in the utterance, and
$L_{2}$ denotes a $L_2$ norm function.

\vspace{-4mm}
\subsection{Secondary Task: Prosody Modeling}
\vspace{-1mm}
The secondary task optimizes a prosody generator to predict the phrase break pattern for each word in the input text, as shown in Fig. \ref{model_fig}(b). We define prosody embedding ($P\!E$) as a vector of five elements, namely \textit{break, non-break, blank, punctuation} and \textit{stop token}, that represents five phrase break patterns. \textit{stop token} denotes the end of an utterance, while \textit{punctuation} refers to any punctuation symbols other than \textit{stop token}.

The word sequence in the input text is first represented by a sequence of word embeddings. We devise a Bidirectional Long Short-Term Memory (BLSTM) as the prosody generator, that takes the word embeddings $W\!E = \{we_1,...,we_t,...,we_T\}$ as input and generates the prosody embedding $P\!E = \{pe_1,...,pe_t,...,pe_T\}$ as output, where $pe_t$ is a 5-dimension embedding vector. Specifically, the forward and backward LSTM reads the word embedding sequence $W\!E$ from both directions. We add a hidden layer on top of the LSTM to detect higher-level feature combinations, and a softmax layer to produce the probability distribution of phrase break patterns $pe_{t}$ for each of the $T$ words.
An element in the embedding vector $pe_{t}= [ p_t[1], ..., p_t[k], ..., p_t[5] ]$, $t \in [1,T] $,  represents the probability of the phrase break label $k$.

The secondary task minimizes the differences between the predicted prosody embedding and the ground truth one-hot vector using the cross-entropy loss $Loss_{pe}=-\sum_{t=1}^{T}\log p_t[k]$,
where $k$ represents the target phrase break pattern.

\vspace{-2mm}
\subsection{Multi-task Learning}

We now have two parallel feature representations of input text, as shown in Fig. \ref{model_fig}(a). $C\!E_o$ is the character representation in the main task and the prosody embedding ($P\!E$) in the secondary task. Prosody embedding serves as an auxiliary input to Tacotron that informs the phrase break information. 
We concatenate $C\!E_o$ and $P\!E$ to form a joint embedding vector as the input for the attention mechanism. In this way, we expect that Tacotron optimizes the voice quality, by also making sure that the phrase break is correct.
As $C\!E_o$ and $P\!E$ have different time resolutions, we upsample the $P\!E$ to align with $C\!E_o$ as shown in Fig. \ref{model_fig}(a).

The total loss function is given as $Loss_{total} = Loss_{wav} + w  \ast Loss_{pe}$, with $w$ as a weight. With the total loss, we expect that the prosody generator learns from both the phrase break annotations and the actual speech utterances to associate acoustic-prosodic patterns with the input text, thus improving the Tacotron spectral generation at run-time inference. The two-task learning strategy is also referred to as the joint training strategy.
\vspace{-2mm}
\section{Experiments}
\label{exp}

\subsection{Databases}
\vspace{-1mm}
\label{data}

{\textbf{Speech Data}}:
We use the TsingHua-Corpus of Speech Synthesis (TH-CoSS) \cite{Cai2007TH} for Chinese. We use a subset of TH-CoSS, denoted as \textit{03FR00}, that contains approximately 9 hours of speech data and 5.6k utterances with 103k words. The speech signals are sampled at 16 kHz and encoded at 16-bit. The Mongolian speech data as in~\cite{li2018end} contains about 17 hours of speech in total. The speech signals are sampled at 22.05 kHz and encoded at 16-bit. For both Chinese and Mongolian, we divide the corpus into training and test sets in a ratio of 4 to 1 in all experiments.

{\textbf{Phrase Break Labels}}:
We use the text transcript of the speech data as the training data of prosody generator. The prosodic phrases of Chinese text, \textit{break} and \textit{non-break},  are manually labelled.  The Mongolian phrase breaks are marked by examining the text and  listening to the speech samples. The \textit{blank}, \textit{punctuation} and \textit{stop token} labels are naturally present in the text. 

{\textbf{Word Embedding}}:
\label{word_emd}
We generate the word embedding $W\!E$ via table look-up. For Chinese, we use the Tencent AI Lab embedding database for Chinese Words and Phrases \cite{N18-2028}. For Mongolian, the pre-trained 200-dimension word embedding reported in \cite{Liu2018Improving} is used.

\vspace{-3mm}
\subsection{Contrastive Systems}

We build three constrastive systems 
 to validate the two ideas in the proposed \textit{MTL-Tacotron}, namely multi-task learning,  and prosody embedding in a comparative study. In all systems, we use Griffin-Lim algorithm \cite{Griffin1984Signal} for waveform generation for rapid turn-around.


1) Traditional \textit{Tacotron}  TTS system as in~\cite{wang2017tacotron}, that doesn't explicitly model prosodic phrasing.

\begin{figure}[t]
  \centering
   \vspace{-3mm}
  \includegraphics[width= 0.42\textwidth]{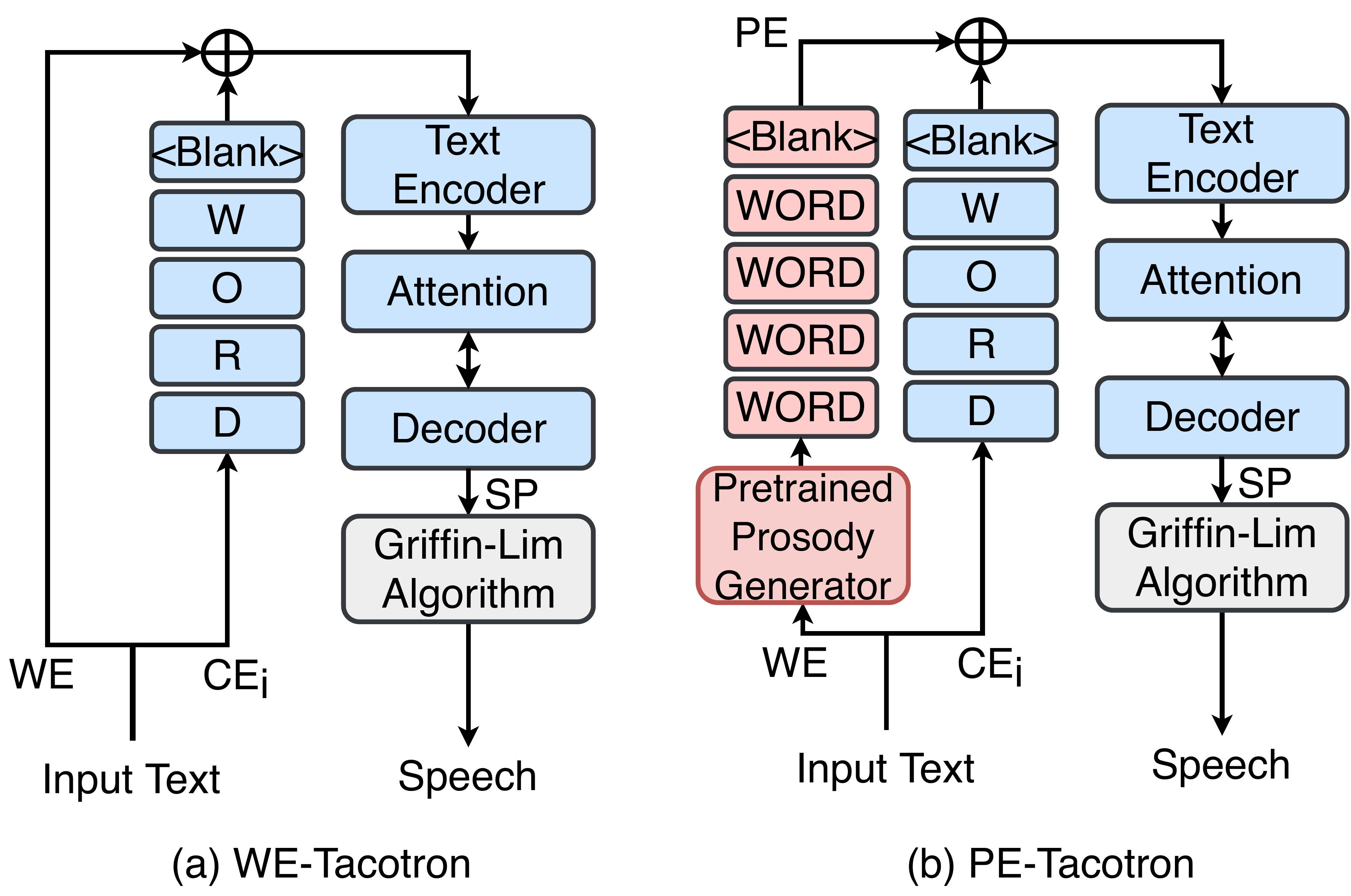}\\
 \vspace{-3mm}
  \caption{Block diagrams of the baseline frameworks (a) WE-Tacotron, and (b) PE-Tacotron. Griffin-Lim algorithm is not involved in training. SP denotes mel-spectrum speech features.}
   \vspace{-2mm}
  \label{baseline_fig}
 
\end{figure}

2) Tacotron augmented with word embedding as in~\cite{chung2019semi}, denoted as \textit{WE-Tacotron} and illustrated in Fig. \ref{baseline_fig} (a). The word embedding informs ~\textit{Tacotron} the word identity and its boundaries, that is shown effective~\cite{chung2019semi}.

3) Tacotron augmented with prosody embedding without multi-task joint training, denoted as \textit{PE-Tacotron} and illustrated in Fig. \ref{baseline_fig} (b). The prosody embeddings are derived from word embedding to encode the prosodic phrasing.


Besides the multi-task learning, \textit{MTL-Tacotron} is also different from both \textit{WE-Tacotron} and \textit{PE-Tacotron} in the way that the text encoder is incorporated in order to facilitate the joint training. 
\textit{PE-Tacotron} and \textit{WE-Tacotron} share similar architecture with \textit{Tacotron} baseline except that \textit{PE-Tacotron} is augmented by prosody embedding, while \textit{WE-Tacotron} is augmented by word embedding. Unlike \textit{MTL-Tacotron}, they incorporate the embeddings that are trained independently of \textit{Tacotron}. They are the contrastive models for \textit{MTL-Tacotron} to show the effect of multi-task learning. 

We note that prosody embedding is derived from word embedding. \textit{MTL-Tacotron} and \textit{PE-Tacotron} are trained to predict the phrase breaks explicitly from word embeddings, while \textit{WE-Tacotron} use the word embeddings directly. Therefore, \textit{WE-Tacotron} serves as the contrastive model for \textit{MTL-Tacotron} and \textit{PE-Tacotron} to show the advantage of the proposed prosody embedding.

\vspace{-2mm}
\subsection{Experimental Setup}
The Chinese text is encoded in \textit{Pinyin} string with tones, and Mongolian text in Latin transliteration. For both languages, we generate 80-channel Mel-spectrum as output. $C\!E_i$ (or $C\!E_o$) and $P\!E$ size are set to 256 and 5 respectively. The number of output frames is controlled by a hyperparameter reduction factor (r), which is set to 5, and the weight $w$ is set to 0.5. We use the Adam optimizer with $\beta_1$ = 0.9, $\beta_2$ = 0.999 and a learning rate of $10^{-3}$ exponentially decaying to $10^{-5}$ starting with 50k steps. We also apply $L_{2}$ regularization with weight $10^{-6}$. All models are trained with a batch size of 32. The final models are trained with 200k steps for all systems.

The prosody generator in the \textit{MTL-Tacotron} is jointly trained with other Tacotron modules. The size of the LSTM layer is set to 200 in both directions for all experiments, the size of hidden layer is set to 50.  The prosody generator in~\textit{PE-Tacotron} has exactly the same configuration as that in~\textit{MTL-Tacotron}, except that it is pre-trained on the training data.

\begin{table}

\centering

\caption {Comparison of phrase break prediction in terms of Precision (P), Recall (R) and F-score (F) for two systems that employ prosody embedding, and mean opinion score (MOS) in listening tests for all systems.}
 \begin{tabular}{cccccc}
 \hline
 \textbf{System}& \textbf{Language} &\textbf{P} &\textbf{R} &\textbf{F} &\textbf{MOS}  \\
\hline 

\multirow{2}{*}{{Tacotron}}  &  Chinese  & NA & NA  & NA & 3.71 {$\pm$} 0.04 \\  \cline{2-6}
 &  Mongolian & NA  & NA  & NA & 3.60 {$\pm$} 0.02\\
\hline 

\multirow{2}{*}{{WE-Tacotron}}  &  Chinese  & NA  & NA & NA  & 3.79 {$\pm$} 0.03\\ \cline{2-6}

 &  Mongolian & NA  & NA  & NA & 3.72 {$\pm$} 0.02\\
\hline 

\multirow{2}{*}{{PE-Tacotron}}  &  Chinese  & 90.01  & 90.88  & 90.68 & 3.86 {$\pm$} 0.02\\ \cline{2-6}

 &  Mongolian & 88.83  & 89.42  & 89.11 & 3.79 {$\pm$} 0.04\\
\hline 

\multirow{2}{*}{\textbf{MTL-Tacotron}}  &  Chinese  & 90.77  & 91.54  & \textbf{91.39} & \textbf{3.91} {$\pm$} 0.01\\ \cline{2-6}

 &  Mongolian & 90.01  & 90.79 & \textbf{90.33}  & \textbf{3.83} {$\pm$} 0.03\\
\hline
\end{tabular}
\label{tab:prfmos}
 \vspace{-2mm}
\end{table}

\vspace{-3mm}
\subsection{Phrase Break Prediction}
We report the phrase break prediction performance of the prosody generator in \textit{MTL-Tacotron} and  \textit{PE-Tacotron} where prosody embedding is used.
As the text in the datasets has already been annotated with prosody labels, it serves as the ground truth for reporting the performance. At run-time inference, the phrase break pattern of a word $we_{t}$ is predicted as $\hat{k} = \arg \max_k { p_t[k]}$. We report the performance in terms of Precision (P), Recall (R) and F-score (F) which is defined as the harmonic mean of the P and R. F values range from 0 to 1, with a higher value indicating better performance. 

As shown in Table \ref{tab:prfmos}, \textit{MTL-Tacotron} clearly outperforms  \textit{PE-Tacotron} in phrase break prediction. By comparing \textit{MTL-Tacotron} and \textit{PE-Tacotron}, we confirm the advantage of joint training over pre-trained prosody embedding.  We expect that \textit{MTL-Tacotron} will reflect the improved phrase break prediction into actual prosodic rendering in speech.

\vspace{-3mm}
\subsection{Subjective Listening Test}
We conduct listening experiments for all systems\footnote{Speech samples in the listening tests: \href{https://ttslr.github.io/SPL2020}{https://ttslr.github.io/SPL2020}}.
20 Chinese and 15 Mongolian speakers participated in the listening tests. Each subject listens to 80 converted utterances of his/her native language. 
\subsubsection{Voice Quality}
We first evaluate the voice quality with mean opinion score (MOS) among these four systems. The listeners rate the quality on a 5-point scale: ``5'' for excellent, ``4'' for good, ``3'' for fair, ``2'' for poor, and ``1'' for bad. In  Table \ref{tab:prfmos}, we observe that \textit{PE-Tacotron} and \textit{MTL-Tacotron} consistently outperform traditional \textit{Tacotron} that doesn't explicitly model prosodic phrasing. The results validate the idea of prosody embedding. Moreover, \textit{MTL-Tacotron} outperforms \textit{WE-Tacotron} and \textit{PE-Tacotron} consistently that confirms the advantage of the proposed joint training.

\subsubsection{Prosodic Embedding vs. Word Embedding}
To confirm the advantage of prosody embedding over word embedding~\cite{chung2019semi}, we further conduct ABX preference tests between pairs of systems. The subjects are asked to choose their preferred utterances in terms of the rhythm and prosody break between a pair of synthesized utterances. The results in Table \ref{tab:AB} suggest that \textit{MTL-Tacotron} system with prosodic phrasing significantly outperforms others in both Chinese and Mongolian experiments.

We also observe that both \textit{MTL-Tacotron} and \textit{PE-Tacotron} outperform \textit{WE-Tacotron} system. As \textit{MTL-Tacotron} and \textit{PE-Tacotron} model the phrase breaks explicitly, the results suggest that modeling phrase breaks explicitly is more effective than using word embedding as a proxy to inform the prosody~\cite{chung2019semi}.
As \textit{MTL-Tacotron} consistently offers superior performance, we are convinced that multi-task learning improves the accuracy of the prosody model over \textit{PE-Tacotron}, thereby generating more accurate prosody embeddings.

\subsubsection{Effect of Text Length}
\begin{figure}[t]
  \centering
  \includegraphics[width=0.48\textwidth]{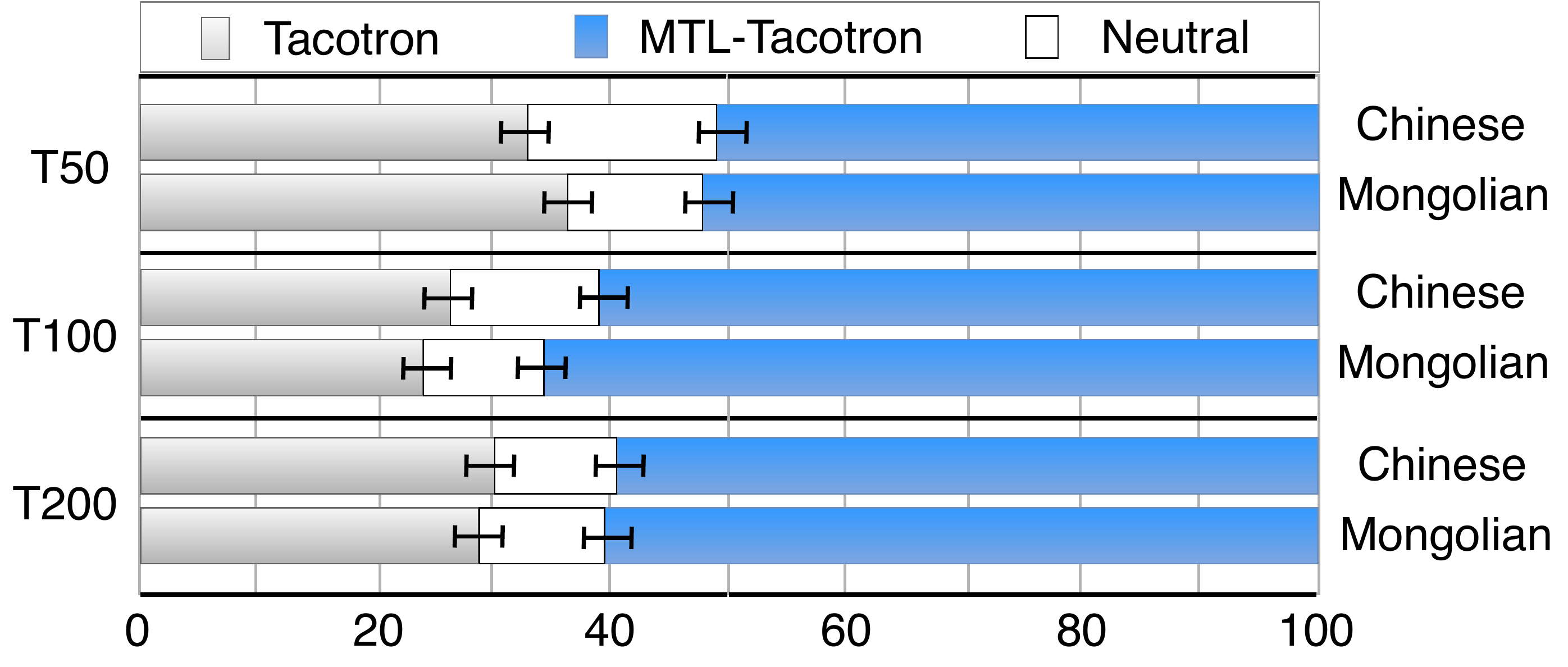}
  \vspace{-4mm}
  \caption{The preference percentage (\%) with 95\% confidence interval between \textit{Tacotron} and \textit{MTL-Tacotron} for various text length.}
  \vspace{-6mm}
  \label{fig:ABtest2}
\end{figure}

We further investigate how the systems perform with regard to the length of input text. By grouping the test sentences by length, we create three subsets: 1) T50 with sentences up to 50 characters; 2) T100 with  sentences of 51 to 100 characters; and 3) T200 with sentences of 101 to 200 characters. 
We select 80 utterances from each group for evaluation of expressiveness, and report the subjective listening test in Fig. \ref{fig:ABtest2}. We observe that \textit{MTL-Tacotron} consistently outperforms the \textit{Tacotron} baseline. It is worth noting that \textit{MTL-Tacotron} performs remarkably well for long sentences for T100 and T200, which is encouraging.

\begin{table}[]
\centering
\caption {The preference percentage (\%) with 95\% confidence interval six competing pairs on common test data.}
\vspace{-2mm}

\begingroup
\renewcommand{\arraystretch}{1} 
\begin{tabular}{p{1.4cm}<{\centering}ccccc}
\hline
\multirow{2}{*}{\textbf{\tabincell{c}{Competing \\ pair}
}} & \multirow{2}{*}{\textbf{Language}} & \multicolumn{3}{c}{\textbf{Preference}(\%)} & \multirow{2}{*}{$\bm{p}$\textbf{-value}} \\ \cline{3-5}
 &  & \textbf{Former} & \textbf{Neutral} & \textbf{Latter} &  \\ \hline
\multirow{2}{*}{ \tabincell{c}{Tacotron \\ vs.\\ WE-Tacotron} } & Chinese & 30.56 & 28.56 & 41.19 &  0.00105\\ [4.5pt] \cline{2-6} 
 & Mongolian & 29.00 &  24.92 & 46.08 & 0.00014 \\ [4.5pt] \hline
\multirow{2}{*}{\begin{tabular}[c]{@{}c@{}}Tacotron\\ vs.\\ PE-Tacotron\end{tabular}} & Chinese & 29.38 & 27.18 & 43.44 & 0.00248 \\[4.5pt] \cline{2-6} 
 & Mongolian & 27.33 & 24.42 & 48.25 & 0.00134 \\[4.5pt] \hline
\multirow{2}{*}{\begin{tabular}[c]{@{}c@{}}Tacotron\\ vs.\\MTL-Tacotron\end{tabular}}& Chinese & 27.44 & 26.25 & 46.31 & 0.00176 \\[4.5pt] \cline{2-6} 
 & Mongolian & 25.91 & 21.01 & 53.08 & 0.00054 \\[4.5pt] \hline
\multirow{2}{*}{\begin{tabular}[c]{@{}c@{}}WE-Tacotron\\ vs.\\ MTL-Tacotron\end{tabular}} & Chinese & 39.56 & 9.75 & 50.69 &0.00392  \\ [4.5pt]\cline{2-6} 
 & Mongolian & 37.42 & 11.41 & 51.17 & 0.00217 \\[4.5pt]\hline
\multirow{2}{*}{\begin{tabular}[c]{@{}c@{}}WE-Tacotron\\ vs.\\ PE-Tacotron\end{tabular}} & Chinese & 38.81 & 9.69 & 51.50 &0.00282  \\ [4.5pt]\cline{2-6} 
 & Mongolian & 37.33 & 12.00 & 50.67 & 0.00153 \\[4.5pt]\hline
\multirow{2}{*}{\begin{tabular}[c]{@{}c@{}}PE-Tacotron\\ vs.\\ MTL-Tacotron\end{tabular}} & Chinese &40.06  & 12.69 & 47.25 & 0.00047 \\ [4.5pt]\cline{2-6} 
 & Mongolian & 41.50 & 9.92 & 48.58 & 0.00318  \\ [4.5pt]\hline
\end{tabular}
\endgroup
\vspace{-2mm}
\label{tab:AB}
\end{table}

\vspace{-3mm}
\section{Conclusions}
\label{con}
We have proposed a novel multi-task Tacotron model to model the prosodic phrasing in speech synthesis, where a word-level prosody generator is introduced as the secondary task. The experiments show that the proposed \textit{MTL-Tacotron} consistently outperforms all contrastive systems. The modeling technique for prosodic phrasing can be easily extended to the modeling of other melodic and rhythmic aspects of speech, such as intonation and stress.




\bibliographystyle{IEEEtran}
{\footnotesize
\bibliography{ref}
}

\end{document}